\documentclass[superscriptaddress,twocolumn,showpacs,aps,prl]{revtex4-1}
\usepackage{amssymb,amsmath}
\usepackage[dvips]{graphicx}
\usepackage[dvipdfm,colorlinks,breaklinks=true,linkcolor=blue,citecolor=blue,urlcolor=blue]{hyperref}

\begin{document}

\title{Criticality and quenched disorder: rare regions vs. Harris criterion}

\author{Thomas Vojta}
\affiliation{Department of Physics, Missouri University of Science and Technology, Rolla, MO 65409, USA}

\author{Jos\'e A. Hoyos}
\affiliation{Instituto de F\'{i}sica de S\~ao Carlos, Universidade de S\~ao Paulo,
C.P. 369, S\~ao Carlos, S\~ao Paulo 13560-970, Brazil}

\begin{abstract}
We employ scaling arguments and optimal fluctuation theory to establish a general
relation between quantum Griffiths singularities and the Harris criterion for quantum phase transitions
in disordered systems. If a clean critical point violates the Harris criterion, it is destabilized by
weak disorder. At the same time, the Griffiths dynamical exponent $z'$
diverges upon approaching the transition, suggesting unconventional critical behavior. In contrast,
if the Harris criterion is fulfilled, power-law Griffiths singularities
can coexist with clean critical behavior but $z'$ saturates at a finite value.
We present applications of our theory to a variety of systems including quantum
spin chains, classical reaction-diffusion systems and metallic magnets; and
we discuss modifications for transitions above the upper critical dimension.
Based on these results we propose a unified classification of phase transitions
in disordered systems.
\end{abstract}

\date{\today}

\maketitle


The effects of quenched disorder on continuous phase transitions have been the subject of intense
theoretical and experimental interest, with applications ranging from condensed matter and atomic physics
to chemistry and biology.
Over the course of this research, two different frameworks for classifying disorder effects have emerged,
one based on the behavior of the average disorder strength under coarse graining, the other focusing
on the properties of rare large disorder fluctuations.

The traditional approach is  based on analyzing the stability of clean critical points against weak disorder
via the Harris criterion \cite{Harris74}.
If the clean correlation length exponent $\nu$ fulfills the inequality $d\nu > 2$
($d$ is the space dimensionality), weak disorder decreases under coarse graining and becomes unimportant on large length
scales. The critical behavior of the disordered system is thus
identical to that of the corresponding clean one. If $d\nu < 2$, weak disorder is relevant, i.e., it increases
under coarse graining, and the transition must change. Motrunich {\it et al}.\
\cite{MMHF00} generalized this idea and and classified critical points
according to the fate of the (average) disorder strength under coarse graining:
If it vanishes on large length scales, the clean critical behavior is unchanged.
If it reaches a nonzero finite value, the critical behavior
remains conventional, but the critical exponents differ from the clean ones. Finally,
if the disorder strength diverges under coarse graining, the transition is governed
by an exotic infinite-randomness critical point.

In recent years, it has become clear that the average disorder strength is insufficient to
characterize disordered critical points. Rare large disorder fluctuations and the corresponding
spatial regions play an important role. Such rare regions can be
locally in one phase while the bulk system is in the other. Griffiths showed
that they cause nonanalyticities, now known as the Griffiths singularities, in the free energy in an
entire parameter region around the transition \cite{Griffiths69,*McCoy69}.
The strength of these singularities can be classified \cite{VojtaSchmalian05}
by comparing the rare-region dimensionality $d_{RR}$ with the lower critical dimension $d_c^-$
of the transition
\footnote{$d_{RR}$ counts the dimensions (incl.\ imaginary time at a QPT) in which the rare region is infinite.
$d_c^-$ is the dimension at and below which the transition does
not exist because the ordered phase is destroyed by fluctuations.}.
If $d_{RR}<d_c^-$, individual rare regions cannot undergo the transition independently.
The resulting weak essential Griffiths singularities are likely unobservable in experiment.
If $d_{RR}>d_c^-$, individual rare regions do order independently;
this destroys the global phase transition by smearing. In the marginal case, $d_{RR}=d_c^-$, rare regions
cannot order, but their slow dynamics leads to enhanced \emph{quantum} Griffiths singularities
characterized by power laws with a nonuniversal dynamical exponent $z'$.

These two classification schemes have been used successfully to analyze a plethora of classical, quantum,
and nonequilibrium transitions (see, e.g., Refs.\ \cite{Vojta06,*Vojta10}). However, they look at
different aspects of the disorder problem which may lead to seemingly incompatible predictions, e.g.,
if Harris' inequality $d\nu>2$ is fulfilled while the rare regions produce strong power-law Griffiths
singularities.

In this Letter, we use scaling and optimal fluctuation theory to
establish a general relation between quantum Griffiths singularities and the Harris criterion:
If disorder is introduced into a system that fulfills Harris' inequality $d\nu>2$, power-law 
Griffiths singularities coexist with clean critical behavior, and the dynamical exponent $z'$ governing the
Griffiths singularities saturates at a finite, disorder-dependent value
at the transition.
In contrast, if Harris' inequality is violated, $z'$ diverges upon approaching the transition, suggesting
strong-randomness or infinite-randomness critical behavior.
In the remainder of this Letter, we sketch the derivation of the results, and we discuss several examples
in quantum spin chains \cite{AshkinTeller43,*KohmotoNijsKadanoff81},
classical reaction-diffusion systems \cite{Hinrichsen00,*Odor04}, random quantum Ising models
\cite{ThillHuse95,Fisher95,*RiegerYoung96} and metallic magnets \cite{Hertz76,*Millis93}.
We also consider modifications for transitions above the upper critical dimension where hyperscaling
is violated, and we present computer simulations illustrating our theory. Finally, we use these results
to propose a refined classification of phase transitions in disordered systems.

Our first example is the Ashkin-Teller model \cite{AshkinTeller43,*KohmotoNijsKadanoff81}
which consists of two coupled transverse-field Ising chains:
\begin{eqnarray}
 H=&-&\sum_{\alpha=1}^2\sum_{i}{\left ( J_i \sigma_{\alpha,i}^z \sigma_{\alpha,i+1}^z + h_i \sigma_{\alpha,i}^x \right )}\nonumber \\
&-&\sum_{i}{\epsilon \left ( J_i \sigma_{1,i}^z \sigma_{1,i+1}^z \sigma_{2,i}^z \sigma_{2,i+1}^z +  h_i \sigma_{1,i}^x \sigma_{2,i}^x\right )},
\label{eq:HAT}
\end{eqnarray}
where  $\sigma^x, \sigma^z$ are Pauli matrices, and $J_i$ and $h_i$ denote the interactions and transverse
fields.
In the clean case, $J_i\equiv J$, $h_i\equiv h$, the system undergoes a quantum phase transition
from a paramagnetic phase to a ferromagnetic (Baxter) phase at $h=J$ for all $\epsilon$ between $-1/\sqrt{2}$ and 1.
The critical exponents vary continuously with $\epsilon$. In particular, the correlation length exponent $\nu$ is
below 2 for  $\epsilon > -1/{2}$ but above 2 for  $\epsilon < -1/2$.
Upon introducing weak (random mass) disorder, the local  distance from criticality $r_i = \ln(h_i/J_i)$ becomes a
random variable. It is governed by a probability distribution $W(r_i)$ which we take to be a binary distribution,
$W(r_i) = p\, \delta(r_i - r_h) + (1-p)\, \delta(r_i - r_l)$ with $r_h>r_l$, for simplicity.

Consider a large spatial region of linear size $L_{RR}$ containing $N\sim L_{RR}^d$ sites.
(We formulate the theory in $d$ dimensions, in our example $d=1$). The effective distance from criticality $r$
of this region is given by the average of its local $r_i$. It has a binomial probability distribution
\begin{equation}
P(r,L_{RR}) = \sum_{n=0}^N \binom N n p^n (1-p)^{N-n} \,\delta \left[r- r_{RR}(N,n) \right] .
\label{eq:binomial}
\end{equation}
with $r_{RR}(N,n)=r_l +\frac n N (r_h - r_l)$.
For large regions of roughly average composition, this binomial
can be approximated by a Gaussian
\begin{equation}
P_G(r,L_{RR}) \sim \exp\left[ - \frac 1 {2b^2}\, L_{RR}^d \,(r-r_{av})^2 \right] ,
\label{eq:gaussian}
\end{equation}
where $r_{av}=p r_h +(1-p) r_l$ is the average distance from criticality and $b^2=p(1-p)(r_h-r_l)^2$
measures the strength of the disorder.
Regions with $r<0$ are locally ferromagnetic
even if the bulk system is still paramagnetic, $r_{av}>0$. They thus constitute the rare regions
responsible for quantum Griffiths singularities.

The low-energy spectrum of a single, locally ordered, rare region is equivalent to that
of two coupled two-level systems. The energy gap $\epsilon$ can be easily estimated in perturbation theory
\cite{SenthilSachdev96}, yielding
\begin{equation}
\epsilon(L_{RR}) = \epsilon_0 \exp \left[ -a L_{RR}^d \right] ,
\label{eq:epsilon}
\end{equation}
with $\epsilon_0 \approx h$. According to finite-size scaling (FSS) \cite{Barber_review83},
the coefficient $a$, which has the dimension of an inverse volume, behaves as $a= a' (-r)^{d\nu}$, with $r$ being the distance of the rare region
from criticality \footnote{As the rare region is embedded in a paramagnetic bulk, it has
fluctuating boundary conditions. However, this should not affect the functional form of the FSS law.}. 
Here, $\nu$ represents the \emph{clean} correlation length exponent unless the rare region is in
the narrow asymptotic critical region.

We now consider a system in the paramagnetic phase, $r_{av}>0$, but close to the phase transition.
The rare-region density of states can be estimated by integrating over all locally ordered regions
\footnote{Thill and Huse used a similar integral in their seminal study of the quantum Ising spin glass \cite{ThillHuse95}.},
\begin{eqnarray}
\rho(\epsilon) \sim \int_{-\infty}^0 dr \int_0^\infty dL_{RR} \, P(r,L_{RR})\, \delta\left[ \epsilon -\epsilon(L_{RR}) \right] .
\label{eq:DOS_integral}
\end{eqnarray}
Using the Gaussian approximation Eq.~(\ref{eq:gaussian}) for the joint distribution $P(r,L_{RR})$, this expression can be easily evaluated.
The $L_{RR}$-integral can be carried out exactly while the remaining integral over $r$ can be performed in saddle-point
approximation in the limit $\epsilon\to 0$. The resulting saddle-point value is
\begin{equation}
r_{sp} = \frac {d\nu}{d\nu-2}\, r_{av} .
\label{eq:r_sp}
\end{equation}
Two cases must be distinguished:

(i) If $d\nu < 2$, $r_{sp}$ is negative and thus within the integration
interval $(-\infty,0)$. Inserting $r_{sp}$ into the integral (\ref{eq:DOS_integral}) gives a power-law Griffiths singularity,
\begin{equation}
\rho(\epsilon) \sim \epsilon^{\lambda-1}= \epsilon^{d/z'-1} ,
\label{eq:power_law_DOS}
\end{equation}
in the density of states. Griffiths singularities in various other quantities
can be calculated from Eq.~(\ref{eq:power_law_DOS}). The nonuniversal Griffiths exponent $\lambda$ varies as
\begin{equation}
\lambda \sim b^{-2} r_{av}^{2-d\nu}
\label{eq:lambda}
\end{equation}
with the global distance from criticality, implying that the Griffiths dynamical
exponent $z'$ diverges as $z' \sim b^2 r_{av}^{d\nu-2}$ upon approaching the transition.

(ii) In the opposite case, $d\nu \ge 2$, the maximum of the exponent in the integrand of
(\ref{eq:DOS_integral}) is attained for $r\to -\infty$.
The density of states is thus dominated by contributions from the far tail of the probability
distribution $P(r,L_{RR})$. Thus, the Gaussian approximation,
Eq.~(\ref{eq:gaussian}), is not justified. Instead, one needs to work with the tail of
the original distribution. For our binomial distribution (\ref{eq:binomial}), the far tail consists
of regions in which all sites have $r_i=r_l$. For these regions,
Eq.~(\ref{eq:binomial}) simplifies to
$P(r,L_{RR}) \sim \exp(-\tilde p L_{RR}^d) \,\delta(r-r_l)$ with $\tilde p =-\ln(1-p)$.
As such compact rare regions remain in the ferromagnetic phase when the bulk
reaches criticality, the coefficient $a$ in Eq.~(\ref{eq:epsilon}) takes some \emph{finite nonzero}
value $a_c$ at the bulk transition point
\footnote{Note that $a_c$ vanishes for zero disorder ($r_h=r_l$) and increases
with increasing disorder strength.}.
Combining $P(r,L_{RR})$ and $\epsilon(L_{RR})$, we again find
a power-law density of states as in Eq.~(\ref{eq:power_law_DOS}). However, the Griffiths
exponent $\lambda$ does not vanish at the global transition point but takes the nonzero value
\begin{equation}
\lambda_c =\tilde p /a_c .
\label{eq:lambda_binary}
\end{equation}
This implies that the dynamical exponent $z'$ does not diverge
upon approaching the transition. Its maximum value $z'_c=d a_c/ \tilde p$
vanishes for zero disorder and increases with increasing disorder strength.

We emphasize that our optimal fluctuation theory describes the disorder scaling
close to the clean critical (fixed) point. Therefore, it correctly
describes the asymptotic critical behavior in the case $d\nu>2$. In contrast, for
$d\nu<2$, it does not hold in the asymptotic critical region
because nontrivial disorder renormalizations beyond the tree-level analysis underlying
Eqs.~(\ref{eq:binomial}) and (\ref{eq:gaussian}) become important
as the disorder strength increases.
To explore the limits of our approach, we can use scaling theory which states
that the clean description breaks down when the
scaling combination $b^2 r^{d\nu-2}$ exceeds a constant of order one
\footnote{This follows from the scale dimension of the disorder strength $b^2$
at the clean critical point being $2/\nu-d$ \cite{Cardy_book96}.}.
Up to a numerical factor, the Griffiths dynamical exponent $z'$ in Eq.\ (\ref{eq:lambda})
equals this scaling combination. It thus reaches a value of
order one \emph{independent of the bare disorder strength} before (\ref{eq:lambda}) breaks down.
The further evolution of $z'$ in the asymptotic critical region is beyond the scope
of our method
\footnote{ A strong-disorder renormalization group predicts $z'$ to diverge
at criticality. However, saturation at a finite value may occur
for weaker disorder \cite{CarlonLajkoIgloi01}.}.

We have thus established our main result: The same inequality that controls the scaling of the average disorder
strength also governs the quantum Griffiths singularities. If the clean correlation length exponent $\nu$ fulfills the inequality
$d\nu>2$, the average disorder strength scales to zero under coarse graining. Moreover, the Griffiths dynamical
exponent at the transition takes a finite value that vanishes in the limit of zero disorder and increases
with increasing disorder strength. This means that, for sufficiently weak disorder,
clean critical behavior coexists with subleading power-law quantum Griffiths singularities.
In contrast, for $d\nu<2$, the average disorder strength grows
under coarse graining, destabilizing the clean critical behavior.
The Griffiths dynamical exponent increases in parallel with the renormalized
disorder strength. Even for arbitrarily weak bare disorder, it reaches a value of order one at the
crossover from the clean to the disordered critical fixed point. In the asymptotic
critical region it either diverges or saturates at a large value.
DMRG calculations \cite{CarlonLajkoIgloi01} of short random Ashkin-Teller chains
for selected $\epsilon$ between $-1$ and $1$ are compatible with these predictions.

We now discuss how general our result is. The optimal fluctuation theory applies as long as three assumptions are fulfilled.
(i) The disorder is of random-mass (random-$T_c$) type with short-range correlations and a bounded
probability distribution. (ii) The gap or characteristic energy of a rare region depends exponentially
on its volume putting the system in class B of the rare region classification of Refs.\ \cite{VojtaSchmalian05,Vojta06}.
(iii) The transition is of non-mean-field type (i.e., below the upper critical dimension $d_c^+$) such that
conventional FSS can be used for the relation between the rare region size and its distance
from criticality, $a=a'(-r)^{d\nu}$. These conditions are fulfilled for a large variety of classical, quantum
and nonequilibrium phase transitions in realistic systems~\cite{Vojta06,*Vojta10}.

However, other important (quantum) phase transitions are above $d_c^+$. How is our theory
modified in this case? For $d>d_c^+$, conventional FSS breaks
down due to dangerously irrelevant variables.
Instead, several systems fulfill a modified
FSS \cite{Brezin82,*KennaLang91} (dubbed ``q-scaling'' \cite{BercheKennaWalter12})
that replaces the scaling combination $rL^{1/\nu}$ with
 $rL^{q/\nu}$ where $q=d/d_c^+$.
Examples include the classical Ising model, directed percolation, and the large-$N$ limits of various
$O(N)$ order-parameter field theories.
In the following, we assume that ``q-scaling'' is fulfilled for $d>d_c^+$.

Repeating the derivation of the rare-region density of states, we find that the
only change is in the relation between the rare region size and its distance
from criticality in Eq.~(\ref{eq:epsilon}). Here, $a=a'(-r)^{d\nu}$ gets replaced by
$a=a'(-r)^{d\nu/q}=a'(-r)^{d_c^+\nu}$. As a result, the average disorder
strength and the power-law Griffiths singularities are controlled by different inequalities:
The disorder strength increases under coarse graining if $d\nu < 2$ \cite{[{See also appendix of }] SBBB09}
while the dynamical exponent $z'$ diverges if $d_c^+\nu < 2$.

Let us apply these ideas to our second example, the nonequilibrium transition in the
contact process \cite{HarrisTE74} which can be mapped to a quantum problem using the Hamiltonian
formalism \cite{ADHR94}.
In this problem, each lattice site is in one of two states, infected or healthy.
The time evolution is a Markov process during which
infected sites heal at rate $\mu$ or infect their neighbors at rate
$\kappa$.
If $\mu \gg \kappa$, healing dominates, and the infection eventually dies out completely.
For $\kappa \gg \mu$, the infection never dies out, leading to a nonzero
steady-state density $\rho$ of infected sites.
These two regimes are separated by a nonequilibrium phase transition in the
directed percolation (DP) \cite{GrassbergerdelaTorre79} universality class. It
has an upper critical dimension of $d_c^+=4$. The FSS above $d_c^+$
is of $q$-scaling type \cite{LuebeckJanssen05}.

\begin{figure}
\includegraphics[width=8.cm]{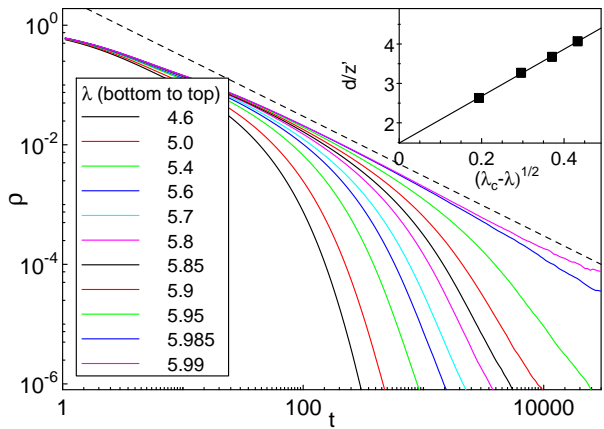}
\caption{(Color online) Density $\rho$ vs. time $t$ for a 5D disordered contact process.
The critical behavior is compatible with the clean mean-field result $\rho \sim t^{-1}$ (dashed line).
The subcritical curves
show Griffiths singularities $\rho \sim t^{-d/z'}$
rather than the exponential decay expected in a clean system.
Inset: extrapolation of the Griffiths exponent $d/z'$ to criticality.}
\label{fig:5dcp}
\end{figure}

In $d=1,2$ and 3, the clean correlation length exponent violates the
inequality $d\nu > 2$. According to the Harris criterion, weak disorder is
relevant. Moreover, Eq.~(\ref{eq:lambda}) predicts the Griffiths dynamical exponent
$z'$ to diverge at the transition. In agreement with these predictions,
infinite-randomness critical behavior has been found in the disordered contact process
 for $d=1, 2$ and 3
\cite{HooyberghsIgloiVanderzande03,VojtaDickison05,*VojtaFarquharMast09,*Vojta12}.
For $d>4$, weak disorder is irrelevant according to the Harris criterion
because $d\nu=d/2 > 2$. Moreover, as $d_c^+\nu =2$, the Griffiths singularities are expected to be dominated
by compact rare regions at the lower bound of the disorder distribution.
Our theory thus predicts that the (weakly) disordered contact process
in $d>4$ features clean critical behavior, accompanied by power-law
Griffiths singularities whose dynamical exponent $z'$ saturates at a finite value
at the transition point. We have performed Monte Carlo simulations of
the disordered five-dimensional contact process \cite{VojtaIgo_unpublished} on lattices with up to $51^5$ sites.
The results (see Fig.\ \ref{fig:5dcp}) are in agreement with these predictions.
Similar behavior has also been observed for the contact process on networks
\cite{JOCM12}.

Our final example is the large-$N$ limit of the quantum Landau-Ginzburg-Wilson (LGW) theory
\begin{equation}
S=\int dx\,dy\ \phi (x)\,\Gamma (x,y)\,\phi (y)+\frac{u}{2N}\int dx\ \phi
^{4}(x)  \label{eq:action}
\end{equation}
in $d$ dimensions. $\phi$ is an $N$-component order parameter, $x\equiv (\mathbf{x},\tau )$ comprises position $\mathbf{x}$ and
imaginary time $\tau $, and $\int dx\equiv \int d\mathbf{x}%
\int_{0}^{1/T}d\tau $. The Fourier transform of the clean inverse propagator $\Gamma (x,y)$
is given by
\begin{equation}
\Gamma (\mathbf{q},\omega _{n})=(r+\mathbf{q}^{2}+\gamma |\omega
_{n}|) .
\end{equation}
This theory describes, e.g., quantum phase transitions in itinerant antiferromagnets
\cite{Hertz76,Millis93} and superconducting nanowires \cite{SachdevWernerTroyer04}.
Its upper critical dimension is $d_c^+=2$.

As the clean correlation length exponent takes the values $\nu=1$ in $d=1$ and $\nu=1/2$
for $d\ge 2$, the Harris criterion $d\nu>2$ is violated for dimensions
$d<4$ but fulfilled for $d>4$. The rare regions in this problem were studied in
Ref.\ \cite{VojtaSchmalian05}, giving a characteristic rare-region energy
$\epsilon = \epsilon_0 \exp[-a' (-r) L_{RR}^d]$ for all dimensions above and below $d_c^+$.
This implies that the FSS above $d_c^+$ is of $q$-scaling type.
As $d_c^+\nu=1$, the inequality $d_c^+\nu >2$ is violated in all dimensions.
For $d<4$, we thus expect weak disorder to be relevant and the Griffiths
dynamical exponent $z'$ to diverge at the transition, in agreement with strong-disorder
renormalization group studies \cite{HoyosKotabageVojta07,*VojtaKotabageHoyos09} that yield
infinite-randomness criticality. In contrast, for $d>4$, our theory predicts $z'$
to diverge even though the Harris criterion is fulfilled,
suggesting that rare regions change
the character of the transition despite the disorder being perturbatively irrelevant.

Note that a similar situation occurs in the transverse-field Ising model
in $d>4$. The Harris criterion is fulfilled but $d_c^+\nu = 3/2 <2$ suggesting a diverging
$z'$ in all dimensions. Interestingly, recent strong-disorder renormalization group calculations
\cite{KovacsIgloi11} show infinite-randomness criticality even for infinite dimensions.

Let us return to the classification of critical points in weakly disordered systems according to the
rare region dimensionality as put forward in Ref.\ \cite{VojtaSchmalian05} and discussed in the
introduction. In the present Letter,
we have studied class B of this classification which contains systems whose rare regions
are right at the lower critical dimension, $d_{RR}=d_c^-$, leading to power-law Griffiths singularities.
According to our results, class B can be subdivided by means of the Harris criterion into class
B1 where clean critical behavior coexists with subleading Griffiths singularities and class B2
featuring strong or infinite-randomness criticality.  Note that this subdivison
applies to non-mean-field transitions. As discussed above, further complications may occur
above $d_c^+$.

In summary, we have established a general relation between the Harris criterion and rare-region
effects in weakly disordered systems. For non-mean-field clean critical points, the scaling of the average disorder strength
under coarse graining and the behavior of the quantum Griffiths singularities are governed by the
same inequality. For $d\nu > 2$, weak disorder is irrelevant,
and the Griffiths dynamical exponent $z'$ remains finite and small at the transition.
If $d\nu < 2$, weak disorder is relevant; and $z'$ increases with the renormalized
disorder strength upon approaching the transition. Above the upper critical dimension,
the situation is more complex. The scaling of the average disorder strength is still governed by the
Harris criterion $d\nu>2$ but the fate of the Griffiths dynamical exponent is controlled by
the inequality $d_c^+\nu >2$. This opens up the exciting possibility that non-perturbative
rare region physics can modify the transition even if the Harris criterion is fulfilled.


This work was supported by the NSF under Grant Nos.\ DMR-1205803
and PHYS-1066293, by Simons Foundation, by FAPESP under Grant No.\ 2013/09850-7, and by CNPq under Grant
Nos.\ 590093/2011-8 and 305261/2012-6. We acknowledge the hospitality of the
Aspen Center for Physics.

\bibliographystyle{apsrev4-1}
\bibliography{../00Bibtex/rareregions}

\begin{thebibliography}{44}%
\makeatletter
\providecommand \@ifxundefined [1]{%
 \@ifx{#1\undefined}
}%
\providecommand \@ifnum [1]{%
 \ifnum #1\expandafter \@firstoftwo
 \else \expandafter \@secondoftwo
 \fi
}%
\providecommand \@ifx [1]{%
 \ifx #1\expandafter \@firstoftwo
 \else \expandafter \@secondoftwo
 \fi
}%
\providecommand \natexlab [1]{#1}%
\providecommand \enquote  [1]{``#1''}%
\providecommand \bibnamefont  [1]{#1}%
\providecommand \bibfnamefont [1]{#1}%
\providecommand \citenamefont [1]{#1}%
\providecommand \href@noop [0]{\@secondoftwo}%
\providecommand \href [0]{\begingroup \@sanitize@url \@href}%
\providecommand \@href[1]{\@@startlink{#1}\@@href}%
\providecommand \@@href[1]{\endgroup#1\@@endlink}%
\providecommand \@sanitize@url [0]{\catcode `\\12\catcode `\$12\catcode
  `\&12\catcode `\#12\catcode `\^12\catcode `\_12\catcode `\%12\relax}%
\providecommand \@@startlink[1]{}%
\providecommand \@@endlink[0]{}%
\providecommand \url  [0]{\begingroup\@sanitize@url \@url }%
\providecommand \@url [1]{\endgroup\@href {#1}{\urlprefix }}%
\providecommand \urlprefix  [0]{URL }%
\providecommand \Eprint [0]{\href }%
\providecommand \doibase [0]{http://dx.doi.org/}%
\providecommand \selectlanguage [0]{\@gobble}%
\providecommand \bibinfo  [0]{\@secondoftwo}%
\providecommand \bibfield  [0]{\@secondoftwo}%
\providecommand \translation [1]{[#1]}%
\providecommand \BibitemOpen [0]{}%
\providecommand \bibitemStop [0]{}%
\providecommand \bibitemNoStop [0]{.\EOS\space}%
\providecommand \EOS [0]{\spacefactor3000\relax}%
\providecommand \BibitemShut  [1]{\csname bibitem#1\endcsname}%
\let\auto@bib@innerbib\@empty
\bibitem [{\citenamefont {Harris}(1974{\natexlab{a}})}]{Harris74}%
  \BibitemOpen
  \bibfield  {author} {\bibinfo {author} {\bibfnamefont {A.~B.}\ \bibnamefont
  {Harris}},\ }\href {\doibase 10.1088/0022-3719/7/9/009} {\bibfield  {journal}
  {\bibinfo  {journal} {J. Phys. C}\ }\textbf {\bibinfo {volume} {7}},\
  \bibinfo {pages} {1671} (\bibinfo {year} {1974}{\natexlab{a}})}\BibitemShut
  {NoStop}%
\bibitem [{\citenamefont {Motrunich}\ \emph {et~al.}(2000)\citenamefont
  {Motrunich}, \citenamefont {Mau}, \citenamefont {Huse},\ and\ \citenamefont
  {Fisher}}]{MMHF00}%
  \BibitemOpen
  \bibfield  {author} {\bibinfo {author} {\bibfnamefont {O.}~\bibnamefont
  {Motrunich}}, \bibinfo {author} {\bibfnamefont {S.~C.}\ \bibnamefont {Mau}},
  \bibinfo {author} {\bibfnamefont {D.~A.}\ \bibnamefont {Huse}}, \ and\
  \bibinfo {author} {\bibfnamefont {D.~S.}\ \bibnamefont {Fisher}},\ }\href
  {\doibase 10.1103/PhysRevB.61.1160} {\bibfield  {journal} {\bibinfo
  {journal} {Phys. Rev. B}\ }\textbf {\bibinfo {volume} {61}},\ \bibinfo
  {pages} {1160} (\bibinfo {year} {2000})}\BibitemShut {NoStop}%
\bibitem [{\citenamefont {Griffiths}(1969)}]{Griffiths69}%
  \BibitemOpen
  \bibfield  {author} {\bibinfo {author} {\bibfnamefont {R.~B.}\ \bibnamefont
  {Griffiths}},\ }\href {\doibase 10.1103/PhysRevLett.23.17} {\bibfield
  {journal} {\bibinfo  {journal} {Phys. Rev. Lett.}\ }\textbf {\bibinfo
  {volume} {23}},\ \bibinfo {pages} {17} (\bibinfo {year} {1969})}\BibitemShut
  {NoStop}%
\bibitem [{\citenamefont {McCoy}(1969)}]{McCoy69}%
  \BibitemOpen
  \bibfield  {author} {\bibinfo {author} {\bibfnamefont {B.~M.}\ \bibnamefont
  {McCoy}},\ }\href {\doibase 10.1103/PhysRevLett.23.383} {\bibfield  {journal}
  {\bibinfo  {journal} {Phys. Rev. Lett.}\ }\textbf {\bibinfo {volume} {23}},\
  \bibinfo {pages} {383} (\bibinfo {year} {1969})}\BibitemShut {NoStop}%
\bibitem [{\citenamefont {Vojta}\ and\ \citenamefont
  {Schmalian}(2005)}]{VojtaSchmalian05}%
  \BibitemOpen
  \bibfield  {author} {\bibinfo {author} {\bibfnamefont {T.}~\bibnamefont
  {Vojta}}\ and\ \bibinfo {author} {\bibfnamefont {J.}~\bibnamefont
  {Schmalian}},\ }\href {\doibase 10.1103/PhysRevB.72.045438} {\bibfield
  {journal} {\bibinfo  {journal} {Phys. Rev. B}\ }\textbf {\bibinfo {volume}
  {72}},\ \bibinfo {pages} {045438} (\bibinfo {year} {2005})}\BibitemShut
  {NoStop}%
\bibitem [{Note1()}]{Note1}%
  \BibitemOpen
  \bibinfo {note} {$d_{RR}$ counts the dimensions (incl.\ imaginary time at a
  QPT) in which the rare region is infinite. $d_c^-$ is the dimension at and
  below which the transition does not exist because the ordered phase is
  destroyed by fluctuations.}\BibitemShut {Stop}%
\bibitem [{\citenamefont {Vojta}(2006)}]{Vojta06}%
  \BibitemOpen
  \bibfield  {author} {\bibinfo {author} {\bibfnamefont {T.}~\bibnamefont
  {Vojta}},\ }\href {\doibase 10.1088/0305-4470/39/22/R01} {\bibfield
  {journal} {\bibinfo  {journal} {J. Phys. A}\ }\textbf {\bibinfo {volume}
  {39}},\ \bibinfo {pages} {R143} (\bibinfo {year} {2006})}\BibitemShut
  {NoStop}%
\bibitem [{\citenamefont {Vojta}(2010)}]{Vojta10}%
  \BibitemOpen
  \bibfield  {author} {\bibinfo {author} {\bibfnamefont {T.}~\bibnamefont
  {Vojta}},\ }\href {\doibase 10.1007/s10909-010-0205-4} {\bibfield  {journal}
  {\bibinfo  {journal} {J. Low Temp. Phys.}\ }\textbf {\bibinfo {volume}
  {161}},\ \bibinfo {pages} {299} (\bibinfo {year} {2010})}\BibitemShut
  {NoStop}%
\bibitem [{\citenamefont {Ashkin}\ and\ \citenamefont
  {Teller}(1943)}]{AshkinTeller43}%
  \BibitemOpen
  \bibfield  {author} {\bibinfo {author} {\bibfnamefont {J.}~\bibnamefont
  {Ashkin}}\ and\ \bibinfo {author} {\bibfnamefont {E.}~\bibnamefont
  {Teller}},\ }\href {\doibase 10.1103/PhysRev.64.178} {\bibfield  {journal}
  {\bibinfo  {journal} {Phys. Rev.}\ }\textbf {\bibinfo {volume} {64}},\
  \bibinfo {pages} {178} (\bibinfo {year} {1943})}\BibitemShut {NoStop}%
\bibitem [{\citenamefont {Kohmoto}\ \emph {et~al.}(1981)\citenamefont
  {Kohmoto}, \citenamefont {den Nijs},\ and\ \citenamefont
  {Kadanoff}}]{KohmotoNijsKadanoff81}%
  \BibitemOpen
  \bibfield  {author} {\bibinfo {author} {\bibfnamefont {M.}~\bibnamefont
  {Kohmoto}}, \bibinfo {author} {\bibfnamefont {M.}~\bibnamefont {den Nijs}}, \
  and\ \bibinfo {author} {\bibfnamefont {L.~P.}\ \bibnamefont {Kadanoff}},\
  }\href {\doibase 10.1103/PhysRevB.24.5229} {\bibfield  {journal} {\bibinfo
  {journal} {Phys. Rev. B}\ }\textbf {\bibinfo {volume} {24}},\ \bibinfo
  {pages} {5229} (\bibinfo {year} {1981})}\BibitemShut {NoStop}%
\bibitem [{\citenamefont {Hinrichsen}(2000)}]{Hinrichsen00}%
  \BibitemOpen
  \bibfield  {author} {\bibinfo {author} {\bibfnamefont {H.}~\bibnamefont
  {Hinrichsen}},\ }\href {\doibase 10.1080/00018730050198152} {\bibfield
  {journal} {\bibinfo  {journal} {Adv. Phys.}\ }\textbf {\bibinfo {volume}
  {49}},\ \bibinfo {pages} {815} (\bibinfo {year} {2000})}\BibitemShut
  {NoStop}%
\bibitem [{\citenamefont {Odor}(2004)}]{Odor04}%
  \BibitemOpen
  \bibfield  {author} {\bibinfo {author} {\bibfnamefont {G.}~\bibnamefont
  {Odor}},\ }\href {\doibase 10.1103/RevModPhys.76.663} {\bibfield  {journal}
  {\bibinfo  {journal} {Rev. Mod. Phys.}\ }\textbf {\bibinfo {volume} {76}},\
  \bibinfo {pages} {663} (\bibinfo {year} {2004})}\BibitemShut {NoStop}%
\bibitem [{\citenamefont {Thill}\ and\ \citenamefont
  {Huse}(1995)}]{ThillHuse95}%
  \BibitemOpen
  \bibfield  {author} {\bibinfo {author} {\bibfnamefont {M.}~\bibnamefont
  {Thill}}\ and\ \bibinfo {author} {\bibfnamefont {D.~A.}\ \bibnamefont
  {Huse}},\ }\href {\doibase 10.1016/0378-4371(94)00247-Q} {\bibfield
  {journal} {\bibinfo  {journal} {Physica A}\ }\textbf {\bibinfo {volume}
  {214}},\ \bibinfo {pages} {321} (\bibinfo {year} {1995})}\BibitemShut
  {NoStop}%
\bibitem [{\citenamefont {Fisher}(1995)}]{Fisher95}%
  \BibitemOpen
  \bibfield  {author} {\bibinfo {author} {\bibfnamefont {D.~S.}\ \bibnamefont
  {Fisher}},\ }\href {\doibase 10.1103/PhysRevB.51.6411} {\bibfield  {journal}
  {\bibinfo  {journal} {Phys. Rev. B}\ }\textbf {\bibinfo {volume} {51}},\
  \bibinfo {pages} {6411} (\bibinfo {year} {1995})}\BibitemShut {NoStop}%
\bibitem [{\citenamefont {Rieger}\ and\ \citenamefont
  {Young}(1996)}]{RiegerYoung96}%
  \BibitemOpen
  \bibfield  {author} {\bibinfo {author} {\bibfnamefont {H.}~\bibnamefont
  {Rieger}}\ and\ \bibinfo {author} {\bibfnamefont {A.~P.}\ \bibnamefont
  {Young}},\ }\href {\doibase 10.1103/PhysRevB.54.3328} {\bibfield  {journal}
  {\bibinfo  {journal} {Phys. Rev. B}\ }\textbf {\bibinfo {volume} {54}},\
  \bibinfo {pages} {3328} (\bibinfo {year} {1996})}\BibitemShut {NoStop}%
\bibitem [{\citenamefont {Hertz}(1976)}]{Hertz76}%
  \BibitemOpen
  \bibfield  {author} {\bibinfo {author} {\bibfnamefont {J.}~\bibnamefont
  {Hertz}},\ }\href {\doibase 10.1103/PhysRevB.14.1165} {\bibfield  {journal}
  {\bibinfo  {journal} {Phys. Rev. B}\ }\textbf {\bibinfo {volume} {14}},\
  \bibinfo {pages} {1165} (\bibinfo {year} {1976})}\BibitemShut {NoStop}%
\bibitem [{\citenamefont {Millis}(1993)}]{Millis93}%
  \BibitemOpen
  \bibfield  {author} {\bibinfo {author} {\bibfnamefont {A.~J.}\ \bibnamefont
  {Millis}},\ }\href {\doibase 10.1103/PhysRevB.48.7183} {\bibfield  {journal}
  {\bibinfo  {journal} {Phys. Rev. B}\ }\textbf {\bibinfo {volume} {48}},\
  \bibinfo {pages} {7183} (\bibinfo {year} {1993})}\BibitemShut {NoStop}%
\bibitem [{\citenamefont {Senthil}\ and\ \citenamefont
  {Sachdev}(1996)}]{SenthilSachdev96}%
  \BibitemOpen
  \bibfield  {author} {\bibinfo {author} {\bibfnamefont {T.}~\bibnamefont
  {Senthil}}\ and\ \bibinfo {author} {\bibfnamefont {S.}~\bibnamefont
  {Sachdev}},\ }\href {\doibase 10.1103/PhysRevLett.77.5292} {\bibfield
  {journal} {\bibinfo  {journal} {Phys. Rev. Lett.}\ }\textbf {\bibinfo
  {volume} {77}},\ \bibinfo {pages} {5292} (\bibinfo {year}
  {1996})}\BibitemShut {NoStop}%
\bibitem [{\citenamefont {Barber}(1983)}]{Barber_review83}%
  \BibitemOpen
  \bibfield  {author} {\bibinfo {author} {\bibfnamefont {M.~N.}\ \bibnamefont
  {Barber}},\ }in\ \href@noop {} {\emph {\bibinfo {booktitle} {Phase
  Transitions and Critical Phenomena}}},\ Vol.~\bibinfo {volume} {8},\ \bibinfo
  {editor} {edited by\ \bibinfo {editor} {\bibfnamefont {C.}~\bibnamefont
  {Domb}}\ and\ \bibinfo {editor} {\bibfnamefont {J.~L.}\ \bibnamefont
  {Lebowitz}}}\ (\bibinfo  {publisher} {Academic},\ \bibinfo {address} {New
  York},\ \bibinfo {year} {1983})\ pp.\ \bibinfo {pages} {145--266}\BibitemShut
  {NoStop}%
\bibitem [{Note2()}]{Note2}%
  \BibitemOpen
  \bibinfo {note} {As the rare region is embedded in a paramagnetic bulk, it
  has fluctuating boundary conditions. However, this should not affect the
  functional form of the FSS law.}\BibitemShut {Stop}%
\bibitem [{Note3()}]{Note3}%
  \BibitemOpen
  \bibinfo {note} {Thill and Huse used a similar integral in their seminal
  study of the quantum Ising spin glass \cite {ThillHuse95}.}\BibitemShut
  {Stop}%
\bibitem [{Note4()}]{Note4}%
  \BibitemOpen
  \bibinfo {note} {Note that $a_c$ vanishes for zero disorder ($r_h=r_l$) and
  increases with increasing disorder strength.}\BibitemShut {Stop}%
\bibitem [{Note5()}]{Note5}%
  \BibitemOpen
  \bibinfo {note} {This follows from the scale dimension of the disorder
  strength $b^2$ at the clean critical point being $2/\nu -d$ \cite
  {Cardy_book96}.}\BibitemShut {Stop}%
\bibitem [{Note6()}]{Note6}%
  \BibitemOpen
  \bibinfo {note} {A strong-disorder renormalization group predicts $z'$ to
  diverge at criticality. However, saturation at a finite value may occur for
  weaker disorder \cite {CarlonLajkoIgloi01}.}\BibitemShut {Stop}%
\bibitem [{\citenamefont {Carlon}\ \emph {et~al.}(2001)\citenamefont {Carlon},
  \citenamefont {Lajk\'o},\ and\ \citenamefont {Igl\'oi}}]{CarlonLajkoIgloi01}%
  \BibitemOpen
  \bibfield  {author} {\bibinfo {author} {\bibfnamefont {E.}~\bibnamefont
  {Carlon}}, \bibinfo {author} {\bibfnamefont {P.}~\bibnamefont {Lajk\'o}}, \
  and\ \bibinfo {author} {\bibfnamefont {F.}~\bibnamefont {Igl\'oi}},\ }\href
  {\doibase 10.1103/PhysRevLett.87.277201} {\bibfield  {journal} {\bibinfo
  {journal} {Phys. Rev. Lett.}\ }\textbf {\bibinfo {volume} {87}},\ \bibinfo
  {pages} {277201} (\bibinfo {year} {2001})}\BibitemShut {NoStop}%
\bibitem [{\citenamefont {Brezin}(1982)}]{Brezin82}%
  \BibitemOpen
  \bibfield  {author} {\bibinfo {author} {\bibfnamefont {E.}~\bibnamefont
  {Brezin}},\ }\href {\doibase 10.1051/jphys:0198200430101500} {\bibfield
  {journal} {\bibinfo  {journal} {J. Phys. (France)}\ }\textbf {\bibinfo
  {volume} {43}},\ \bibinfo {pages} {15} (\bibinfo {year} {1982})}\BibitemShut
  {NoStop}%
\bibitem [{\citenamefont {Kenna}\ and\ \citenamefont
  {Lang}(1991)}]{KennaLang91}%
  \BibitemOpen
  \bibfield  {author} {\bibinfo {author} {\bibfnamefont {R.}~\bibnamefont
  {Kenna}}\ and\ \bibinfo {author} {\bibfnamefont {C.~B.}\ \bibnamefont
  {Lang}},\ }\href {\doibase http://dx.doi.org/10.1016/0370-2693(91)90367-Y}
  {\bibfield  {journal} {\bibinfo  {journal} {Phys. Lett. B}\ }\textbf
  {\bibinfo {volume} {264}},\ \bibinfo {pages} {396} (\bibinfo {year}
  {1991})}\BibitemShut {NoStop}%
\bibitem [{\citenamefont {Berche}\ \emph {et~al.}(2012)\citenamefont {Berche},
  \citenamefont {Kenna},\ and\ \citenamefont {Walter}}]{BercheKennaWalter12}%
  \BibitemOpen
  \bibfield  {author} {\bibinfo {author} {\bibfnamefont {B.}~\bibnamefont
  {Berche}}, \bibinfo {author} {\bibfnamefont {R.}~\bibnamefont {Kenna}}, \
  and\ \bibinfo {author} {\bibfnamefont {J.-C.}\ \bibnamefont {Walter}},\
  }\href {\doibase http://dx.doi.org/10.1016/j.nuclphysb.2012.07.021}
  {\bibfield  {journal} {\bibinfo  {journal} {Nucl. Phys. B}\ }\textbf
  {\bibinfo {volume} {865}},\ \bibinfo {pages} {115} (\bibinfo {year}
  {2012})}\BibitemShut {NoStop}%
\bibitem [{\citenamefont {Sarlat}\ \emph {et~al.}(2009)\citenamefont {Sarlat},
  \citenamefont {Billoire}, \citenamefont {Biroli},\ and\ \citenamefont
  {Bouchaud}}]{SBBB09}%
  \BibitemOpen
  \bibfield  {author} {\bibinfo {author} {\bibfnamefont {T.}~\bibnamefont
  {Sarlat}}, \bibinfo {author} {\bibfnamefont {A.}~\bibnamefont {Billoire}},
  \bibinfo {author} {\bibfnamefont {G.}~\bibnamefont {Biroli}}, \ and\ \bibinfo
  {author} {\bibfnamefont {J.-P.}\ \bibnamefont {Bouchaud}},\ }\href {\doibase
  10.1088/1742-5468/2009/08/P08014} {\bibfield  {journal} {\bibinfo  {journal}
  {J. Stat. Mech.}\ }\textbf {\bibinfo {volume} {2009}},\ \bibinfo {pages}
  {P08014} (\bibinfo {year} {2009})}\BibitemShut {NoStop}%
\bibitem [{\citenamefont {Harris}(1974{\natexlab{b}})}]{HarrisTE74}%
  \BibitemOpen
  \bibfield  {author} {\bibinfo {author} {\bibfnamefont {T.~E.}\ \bibnamefont
  {Harris}},\ }\href {\doibase doi:10.1214/aop/1176996493} {\bibfield
  {journal} {\bibinfo  {journal} {Ann. Prob.}\ }\textbf {\bibinfo {volume}
  {2}},\ \bibinfo {pages} {969} (\bibinfo {year}
  {1974}{\natexlab{b}})}\BibitemShut {NoStop}%
\bibitem [{\citenamefont {Alcaraz}\ \emph {et~al.}(1994)\citenamefont
  {Alcaraz}, \citenamefont {Droz}, \citenamefont {Henkel},\ and\ \citenamefont
  {Rittenberg}}]{ADHR94}%
  \BibitemOpen
  \bibfield  {author} {\bibinfo {author} {\bibfnamefont {F.~C.}\ \bibnamefont
  {Alcaraz}}, \bibinfo {author} {\bibfnamefont {M.}~\bibnamefont {Droz}},
  \bibinfo {author} {\bibfnamefont {M.}~\bibnamefont {Henkel}}, \ and\ \bibinfo
  {author} {\bibfnamefont {V.}~\bibnamefont {Rittenberg}},\ }\href@noop {}
  {\bibfield  {journal} {\bibinfo  {journal} {Ann. Phys. (NY)}\ }\textbf
  {\bibinfo {volume} {230}},\ \bibinfo {pages} {250} (\bibinfo {year}
  {1994})}\BibitemShut {NoStop}%
\bibitem [{\citenamefont {Grassberger}\ and\ \citenamefont {de~la
  Torre}(1979)}]{GrassbergerdelaTorre79}%
  \BibitemOpen
  \bibfield  {author} {\bibinfo {author} {\bibfnamefont {P.}~\bibnamefont
  {Grassberger}}\ and\ \bibinfo {author} {\bibfnamefont {A.}~\bibnamefont
  {de~la Torre}},\ }\href {\doibase 10.1016/0003-4916(79)90207-0} {\bibfield
  {journal} {\bibinfo  {journal} {Ann. Phys. (NY)}\ }\textbf {\bibinfo {volume}
  {122}},\ \bibinfo {pages} {373} (\bibinfo {year} {1979})}\BibitemShut
  {NoStop}%
\bibitem [{\citenamefont {L\"ubeck}\ and\ \citenamefont
  {Janssen}(2005)}]{LuebeckJanssen05}%
  \BibitemOpen
  \bibfield  {author} {\bibinfo {author} {\bibfnamefont {S.}~\bibnamefont
  {L\"ubeck}}\ and\ \bibinfo {author} {\bibfnamefont {H.-K.}\ \bibnamefont
  {Janssen}},\ }\href {\doibase 10.1103/PhysRevE.72.016119} {\bibfield
  {journal} {\bibinfo  {journal} {Phys. Rev. E}\ }\textbf {\bibinfo {volume}
  {72}},\ \bibinfo {pages} {016119} (\bibinfo {year} {2005})}\BibitemShut
  {NoStop}%
\bibitem [{\citenamefont {Hooyberghs}\ \emph {et~al.}(2003)\citenamefont
  {Hooyberghs}, \citenamefont {Igl\'oi},\ and\ \citenamefont
  {Vanderzande}}]{HooyberghsIgloiVanderzande03}%
  \BibitemOpen
  \bibfield  {author} {\bibinfo {author} {\bibfnamefont {J.}~\bibnamefont
  {Hooyberghs}}, \bibinfo {author} {\bibfnamefont {F.}~\bibnamefont {Igl\'oi}},
  \ and\ \bibinfo {author} {\bibfnamefont {C.}~\bibnamefont {Vanderzande}},\
  }\href {\doibase 10.1103/PhysRevLett.90.100601} {\bibfield  {journal}
  {\bibinfo  {journal} {Phys. Rev. Lett.}\ }\textbf {\bibinfo {volume} {90}},\
  \bibinfo {pages} {100601} (\bibinfo {year} {2003})}\BibitemShut {NoStop}%
\bibitem [{\citenamefont {Vojta}\ and\ \citenamefont
  {Dickison}(2005)}]{VojtaDickison05}%
  \BibitemOpen
  \bibfield  {author} {\bibinfo {author} {\bibfnamefont {T.}~\bibnamefont
  {Vojta}}\ and\ \bibinfo {author} {\bibfnamefont {M.}~\bibnamefont
  {Dickison}},\ }\href {\doibase 10.1103/PhysRevE.72.036126} {\bibfield
  {journal} {\bibinfo  {journal} {Phys. Rev. E}\ }\textbf {\bibinfo {volume}
  {72}},\ \bibinfo {pages} {036126} (\bibinfo {year} {2005})}\BibitemShut
  {NoStop}%
\bibitem [{\citenamefont {Vojta}\ \emph
  {et~al.}(2009{\natexlab{a}})\citenamefont {Vojta}, \citenamefont {Farquhar},\
  and\ \citenamefont {Mast}}]{VojtaFarquharMast09}%
  \BibitemOpen
  \bibfield  {author} {\bibinfo {author} {\bibfnamefont {T.}~\bibnamefont
  {Vojta}}, \bibinfo {author} {\bibfnamefont {A.}~\bibnamefont {Farquhar}}, \
  and\ \bibinfo {author} {\bibfnamefont {J.}~\bibnamefont {Mast}},\ }\href
  {\doibase 10.1103/PhysRevE.79.011111} {\bibfield  {journal} {\bibinfo
  {journal} {Phys. Rev. E}\ }\textbf {\bibinfo {volume} {79}},\ \bibinfo
  {pages} {011111} (\bibinfo {year} {2009}{\natexlab{a}})}\BibitemShut
  {NoStop}%
\bibitem [{\citenamefont {Vojta}(2012)}]{Vojta12}%
  \BibitemOpen
  \bibfield  {author} {\bibinfo {author} {\bibfnamefont {T.}~\bibnamefont
  {Vojta}},\ }\href {\doibase 10.1103/PhysRevE.86.051137} {\bibfield  {journal}
  {\bibinfo  {journal} {Phys. Rev. E}\ }\textbf {\bibinfo {volume} {86}},\
  \bibinfo {pages} {051137} (\bibinfo {year} {2012})}\BibitemShut {NoStop}%
\bibitem [{\citenamefont {Vojta}\ and\ \citenamefont
  {Igo}()}]{VojtaIgo_unpublished}%
  \BibitemOpen
  \bibfield  {author} {\bibinfo {author} {\bibfnamefont {T.}~\bibnamefont
  {Vojta}}\ and\ \bibinfo {author} {\bibfnamefont {J.}~\bibnamefont {Igo}},\
  }\href@noop {} {}\bibinfo {note} {Unpublished}\BibitemShut {NoStop}%
\bibitem [{\citenamefont {Juh\'asz}\ \emph {et~al.}(2012)\citenamefont
  {Juh\'asz}, \citenamefont {\'Odor}, \citenamefont {Castellano},\ and\
  \citenamefont {Mu\~noz}}]{JOCM12}%
  \BibitemOpen
  \bibfield  {author} {\bibinfo {author} {\bibfnamefont {R.}~\bibnamefont
  {Juh\'asz}}, \bibinfo {author} {\bibfnamefont {G.}~\bibnamefont {\'Odor}},
  \bibinfo {author} {\bibfnamefont {C.}~\bibnamefont {Castellano}}, \ and\
  \bibinfo {author} {\bibfnamefont {M.~A.}\ \bibnamefont {Mu\~noz}},\ }\href
  {\doibase 10.1103/PhysRevE.85.066125} {\bibfield  {journal} {\bibinfo
  {journal} {Phys. Rev. E}\ }\textbf {\bibinfo {volume} {85}},\ \bibinfo
  {pages} {066125} (\bibinfo {year} {2012})}\BibitemShut {NoStop}%
\bibitem [{\citenamefont {Sachdev}\ \emph {et~al.}(2004)\citenamefont
  {Sachdev}, \citenamefont {Werner},\ and\ \citenamefont
  {Troyer}}]{SachdevWernerTroyer04}%
  \BibitemOpen
  \bibfield  {author} {\bibinfo {author} {\bibfnamefont {S.}~\bibnamefont
  {Sachdev}}, \bibinfo {author} {\bibfnamefont {P.}~\bibnamefont {Werner}}, \
  and\ \bibinfo {author} {\bibfnamefont {M.}~\bibnamefont {Troyer}},\ }\href
  {\doibase 10.1103/PhysRevLett.92.237003} {\bibfield  {journal} {\bibinfo
  {journal} {Phys. Rev. Lett.}\ }\textbf {\bibinfo {volume} {92}},\ \bibinfo
  {pages} {237003} (\bibinfo {year} {2004})}\BibitemShut {NoStop}%
\bibitem [{\citenamefont {Hoyos}\ \emph {et~al.}(2007)\citenamefont {Hoyos},
  \citenamefont {Kotabage},\ and\ \citenamefont
  {Vojta}}]{HoyosKotabageVojta07}%
  \BibitemOpen
  \bibfield  {author} {\bibinfo {author} {\bibfnamefont {J.~A.}\ \bibnamefont
  {Hoyos}}, \bibinfo {author} {\bibfnamefont {C.}~\bibnamefont {Kotabage}}, \
  and\ \bibinfo {author} {\bibfnamefont {T.}~\bibnamefont {Vojta}},\ }\href
  {\doibase 10.1103/PhysRevLett.99.230601} {\bibfield  {journal} {\bibinfo
  {journal} {Phys. Rev. Lett.}\ }\textbf {\bibinfo {volume} {99}},\ \bibinfo
  {pages} {230601} (\bibinfo {year} {2007})}\BibitemShut {NoStop}%
\bibitem [{\citenamefont {Vojta}\ \emph
  {et~al.}(2009{\natexlab{b}})\citenamefont {Vojta}, \citenamefont {Kotabage},\
  and\ \citenamefont {Hoyos}}]{VojtaKotabageHoyos09}%
  \BibitemOpen
  \bibfield  {author} {\bibinfo {author} {\bibfnamefont {T.}~\bibnamefont
  {Vojta}}, \bibinfo {author} {\bibfnamefont {C.}~\bibnamefont {Kotabage}}, \
  and\ \bibinfo {author} {\bibfnamefont {J.~A.}\ \bibnamefont {Hoyos}},\ }\href
  {\doibase 10.1103/PhysRevB.79.024401} {\bibfield  {journal} {\bibinfo
  {journal} {Phys. Rev. B}\ }\textbf {\bibinfo {volume} {79}},\ \bibinfo
  {pages} {024401} (\bibinfo {year} {2009}{\natexlab{b}})}\BibitemShut
  {NoStop}%
\bibitem [{\citenamefont {Kov\'acs}\ and\ \citenamefont
  {Igl\'oi}(2011)}]{KovacsIgloi11}%
  \BibitemOpen
  \bibfield  {author} {\bibinfo {author} {\bibfnamefont {I.~A.}\ \bibnamefont
  {Kov\'acs}}\ and\ \bibinfo {author} {\bibfnamefont {F.}~\bibnamefont
  {Igl\'oi}},\ }\href {\doibase 10.1103/PhysRevB.83.174207} {\bibfield
  {journal} {\bibinfo  {journal} {Phys. Rev. B}\ }\textbf {\bibinfo {volume}
  {83}},\ \bibinfo {pages} {174207} (\bibinfo {year} {2011})}\BibitemShut
  {NoStop}%
\bibitem [{\citenamefont {Cardy}(1996)}]{Cardy_book96}%
  \BibitemOpen
  \bibfield  {author} {\bibinfo {author} {\bibfnamefont {J.}~\bibnamefont
  {Cardy}},\ }\href@noop {} {\emph {\bibinfo {title} {Scaling and
  renormalization in statistical physics}}}\ (\bibinfo  {publisher} {Cambridge
  University Press},\ \bibinfo {address} {Cambridge},\ \bibinfo {year}
  {1996})\BibitemShut {NoStop}%
\end{thebibliography}%

\end{document}